\documentclass{article}
\usepackage{ijcai24}

\usepackage{times}
\usepackage{soul}
\usepackage{url}
\usepackage[utf8]{inputenc}
\usepackage[small]{caption}
\usepackage{graphicx}
\usepackage{subcaption}
\usepackage{amsmath}
\usepackage{amsthm}
\usepackage{amssymb}
\usepackage{booktabs}
\usepackage{algorithm}
\usepackage{algorithmic}
\usepackage[switch]{lineno}
\usepackage{stfloats}
\usepackage{booktabs}
\usepackage{tabularx}
\usepackage{multirow}
\usepackage{color}
\usepackage{amssymb}
\usepackage{makecell} 
\usepackage[most]{tcolorbox}

\definecolor{hotpink}{RGB}{0, 20, 115}
\usepackage[pagebackref=true,breaklinks=true,letterpaper=true,colorlinks,bookmarks=false,citecolor=hotpink]{hyperref}

\title{BATON: Aligning Text-to-Audio Model with Human Preference Feedback}

\author{
Huan Liao$^1$\footnotemark[2]
\and
Haonan Han$^1$\footnotemark[2]\and
Kai Yang$^1$\and
Tianjiao Du$^1$\and
Rui Yang$^1$\and
Zunnan Xu$^1$\\
Qinmei Xu$^1$\and
Jingquan Liu$^1$\and
Jiasheng Lu$^2$\footnotemark[1]\and
Xiu Li$^1$\footnotemark[1]
\affiliations
$^1$Tsinghua University\and
$^2$Huawei Technologies Co., Ltd.\\
\emails
}

\begin{document}

\maketitle

\def\thefootnote{\dag}\footnotetext{Equal contribution}
\def\thefootnote{*}\footnotetext{Corresponding author}
\def\thefootnote{\arabic{footnote}}

\begin{abstract}
    With the development of AI-Generated Content (AIGC), text-to-audio models are gaining widespread attention. However, it is challenging for these models to generate audio aligned with human preference due to the inherent information density of natural language and limited model understanding ability. To alleviate this issue, we formulate the \textbf{BATON}, a framework designed to enhance the alignment between generated audio and text prompt using human preference feedback. Our BATON comprises three key stages: Firstly, we curated a dataset containing both prompts and the corresponding generated audio, which was then annotated based on human feedback. Secondly, we introduced a reward model using the constructed dataset, which can mimic human preference by assigning rewards to input text-audio pairs. Finally, we employed the reward model to fine-tune an off-the-shelf text-to-audio model. The experiment results demonstrate that our BATON can significantly improve the generation quality of the original text-to-audio models, concerning audio integrity, temporal relationship, and alignment with human preference. Project page is available at \url{https://baton2024.github.io}.

\end{abstract}
\section{Introduction}

\begin{figure}[t]
  \centering
  \includegraphics[width=\linewidth]{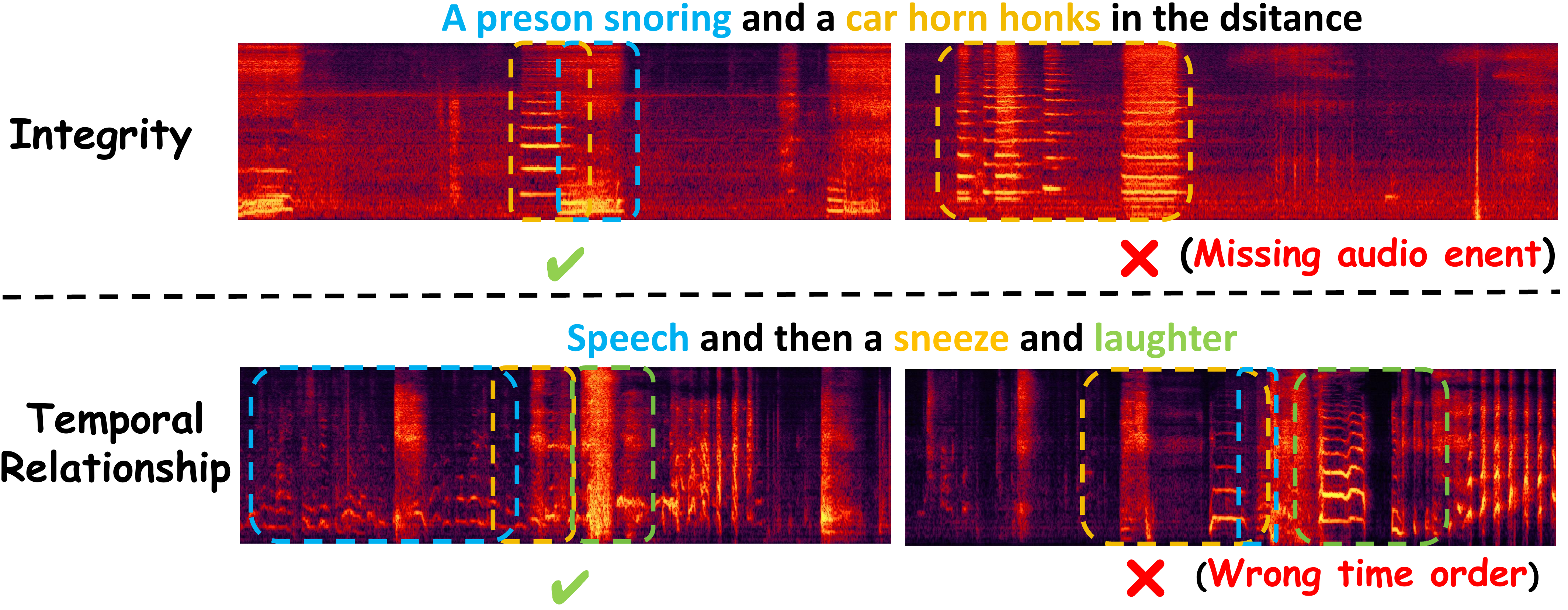}
  \caption{Showcases of two-label and three-label audio samples, with the left indicating alignment and the right indicating misalignment with prompts.} 
  \label{fig:intro-showcase}
\end{figure}

Text-to-audio (TTA) generation is a promising application that concentrates on synthesizing diverse audio from text prompts. Recent advances in diffusion-based generative models, such as AudioLDM~\cite{liu2023audioldm}, Make-An-Audio~\cite{huang2023make}, and TANGO~\cite{ghosal2023tango}, have significantly facilitated audio generation. These models employ latent diffusion models~\cite{rombach2022highldm} to create high-fidelity audio based on textual content, surpassing the performance of previous state-of-the-art TTA models. Nevertheless, there remains a bias between input text and generated audio, caused by the inherent information density of natural language and limited model ability. As shown in Figure~\ref{fig:intro-showcase}, the upper example shows low integrity due to a missed audio event, and the lower example exhibits an incorrect temporal relationship because of a wrong time order. 

Recent studies have made efforts to mitigate this bias. For instance, Make-An-Audio2~\cite{huang2023makeanaudio} utilized large language model to generate structured captions, addressing issues of poor temporal consistency. Re-AudioLDM~\cite{yuan2024retrievalaugmented} tackled the missing generation of rare audio classes through retrieval augmentation. Auffusion~\cite{xue2024auffusion} explored the capabilities of different text encoders to capture fine-grained information such as temporal order and fine-grained events. However, none of these approaches leverage human preference feedback, extensively applied in language~\cite{ouyang2022training,openai2023gpt4} and image generation~\cite{lee2023aligning}, to address the aforementioned issue. Therefore, we aim to explore a new perspective: \textit{Can effective alignment between text prompts and generated audio be achieved through human preference feedback in text-to-audio generation?}

To investigate this, we propose a new framework named BATON, designed for aligning TTA models using human feedback. To the best of our knowledge, this is the first work to improve the alignment of TTA models through human preference feedback. The BATON involves three crucial steps, as depicted in Figure~\ref{fig:overview}: 
\textbf{(1)} We generate initial text prompts using GPT-4~\cite{openai2023gpt4}, conditioned on audio labels obtained from AudioCaps~\cite{Kim_Kim_Lee_Kim_2019} and a self-build conjunction list, as it has proven to be a reliable annotator~\cite{yang2023gpt4tools}. Next, audio is generated for each text prompt using an advanced TTA model, resulting in text-audio pairs. Notably, the generated multi-event audio emphasize the integrity and temporal relationships, aspects that pose challenges for existing TTA models~\cite{xue2024auffusion,liu2023audioldm,ghosal2023tango}. Then, we collect binary human feedback for these pairs. 
\textbf{(2)} We introduce and train an audio reward model, building upon the human-annotated dataset. This model is tailored to predict human feedback based on the provided textual input and corresponding audio. 
\textbf{(3)} We fine-tune an off-the-shelf TTA model, TANGO~\cite{ghosal2023tango}, through reward-weighted likelihood maximization to enhance the alignment of the TTA model with human preference, specifically focusing on the alignment between input text and generated audio. Here, the original loss from TANGO~\cite{ghosal2023tango} is also incorporated as a regularization mechanism, preventing the model from overfitting to the annotated dataset.

We conducted extensive experiments to illustrate that BATON is capable of achieving significant improvements in text-audio alignment. Specifically, BATON demonstrates gains of $+2.3\%$ and $+6.0\%$ in CLAP scores for integrity and temporal relationship tasks, respectively. When evaluated by human annotators, BATON achieves a MOS-Q of $4.55$ for the integrity task, surpassing the original model's score of $4.05$, and attains a MOS-F of $4.41$ for the temporal relationship task, outperforming the original model by $0.58$ MOS-F.

In summary, our main contributions are as follows:
\begin{itemize}
\item We generate a dataset with $4.8$K text-audio pairs across 200 audio event categories, in which $2.7$K samples are annotated by human annotators.

\item Using the constructed dataset, we train an audio reward model to predict the TTA alignment score reflecting human preference. Our experiments demonstrate that scores predicted by the audio reward model closely correlate with human preference.

\item The audio reward models are utilized to enhance an off-the-shelf TTA model in terms of the integrity and temporal relationship of audio events. Detailed experiments provide strong support for its effectiveness.
\end{itemize}

\section{Related Works}

\subsection{Text-to-audio generative models}
The TTA generative models are designed to generate audio that is semantically consistent with text prompt.
Currently, their predominant architecture is founded upon the latent diffusion model~\cite{rombach2022highldm,ho2020ddpm}.
These models engage in a process of noise injection and denoising within the latent space, which is obtained by encoding audio features. 
And the generated audio representation undergoes transformation into a waveform through a vocoder~\cite{kong2020hifi}.

Diffsound~\cite{yang2023diffsound} employs a discrete diffusion model for audio generation from text, operating within the latent space obtained by quantized VAE(VQ-VAE)~\cite{van2017vqvae} trained on mel-spectrograms. 
On the other hand, Audiogen~\cite{kreuk2022audiogen} is built upon a VQ-VAE trained on raw waveform and utilizes an autoregressive language model for the generation of audio guided by text.
For the continuous latent space, AudioLDM~\cite{liu2023audioldm} utilizes audio features obtained from contrastive language-audio pretraining (CLAP)~\cite{wu2023largeclap} as training conditions and guides the generation of audio through textual features encoded by CLAP text encoder, which effectively compensates for the scarcity of paired data.
TANGO~\cite{ghosal2023tango} harnesses the remarkable text representation capability of the pretrained Large Language Model (LLM), achieving superior performance within the same latent space as AudioLDM, with limited paired data.

However, challenges remain for generating audio that matches the details of prompt and human perception. Some recent studies have tried to handle this problem in terms of prompt processing and model components design. Make-An-Audio2 \cite{huang2023makeanaudio} employs LLM to generate structured captions to tackle poor temporal consistency. Yuan \cite{yuan2024retrievalaugmented} solves missing generation of rare audio classes with retrieval augmentation. Guo \cite{guo2023audio} uses timestamps condition for controllable audio generation. Auffusion \cite{xue2024auffusion} explores the ability of different text encoders for capturing fine-grained information like temporal order and fine-grained events. 

\subsection{Fine-tuning with Human Preference Feedback}
Fine-tuning guided by human preference feedback, stands as an indispensable component within the Reinforcement Learning from Human Feedback (RLHF). This approach proves particularly vital in machine learning, especially when dealing with intricate or ambiguous objectives. Its efficacy spans diverse applications, ranging from gaming, as demonstrated in Atari \cite{christiano2017deep}, to more intricate tasks in robotics \cite{2019Fine,casper2023open}, where it significantly enhances the agent's success rate in completing tasks.

The integration of RLHF into the development of large language models (LLMs) signifies a noteworthy milestone in the field. Notable models such as OpenAI's GPT-4 \cite{openai2023gpt4}, Anthropic's Claude \cite{claude}, Google's Bard \cite{Google}, and Meta's Llama 2-Chat \cite{touvron2023llama} leverage this approach to improve their performance and relevance. Collecting human judgments on response quality is often more feasible than obtaining expert demonstrations. Subsequent works have fine-tuned LLMs using datasets reflecting human preferences, resulting in enhanced proficiency in translation \cite{kreutzer-etal-2018-reliability}, summarization \cite{2020Learning}, story-telling \cite{2019Fine}, and instruction-following \cite{ouyang2022training}. Additionally, RLHF has been applied to train language models for various objectives \cite{2015Sequence,2018Learning}. Presently, RLHF has been employed to fine-tune diffusion models, contributing to improved image equality, text-image alignment, and image aesthetic scores \cite{black2024training,dong2023raft,yang2023using}.

Similar to the RLHF method mentioned above, we also trained a reward model for evaluating the quality of audio and applied it to fine-tuning the pre-trained model. Differently, our approach does not use reinforcement learning algorithms to update the original model. Instead, we used the reward model as a weighting factor to increase the probability of audio with higher rewards.

\begin{figure*}[t]
  \centering
  \hspace*{-3.5mm}
  \includegraphics[width=516pt]
  {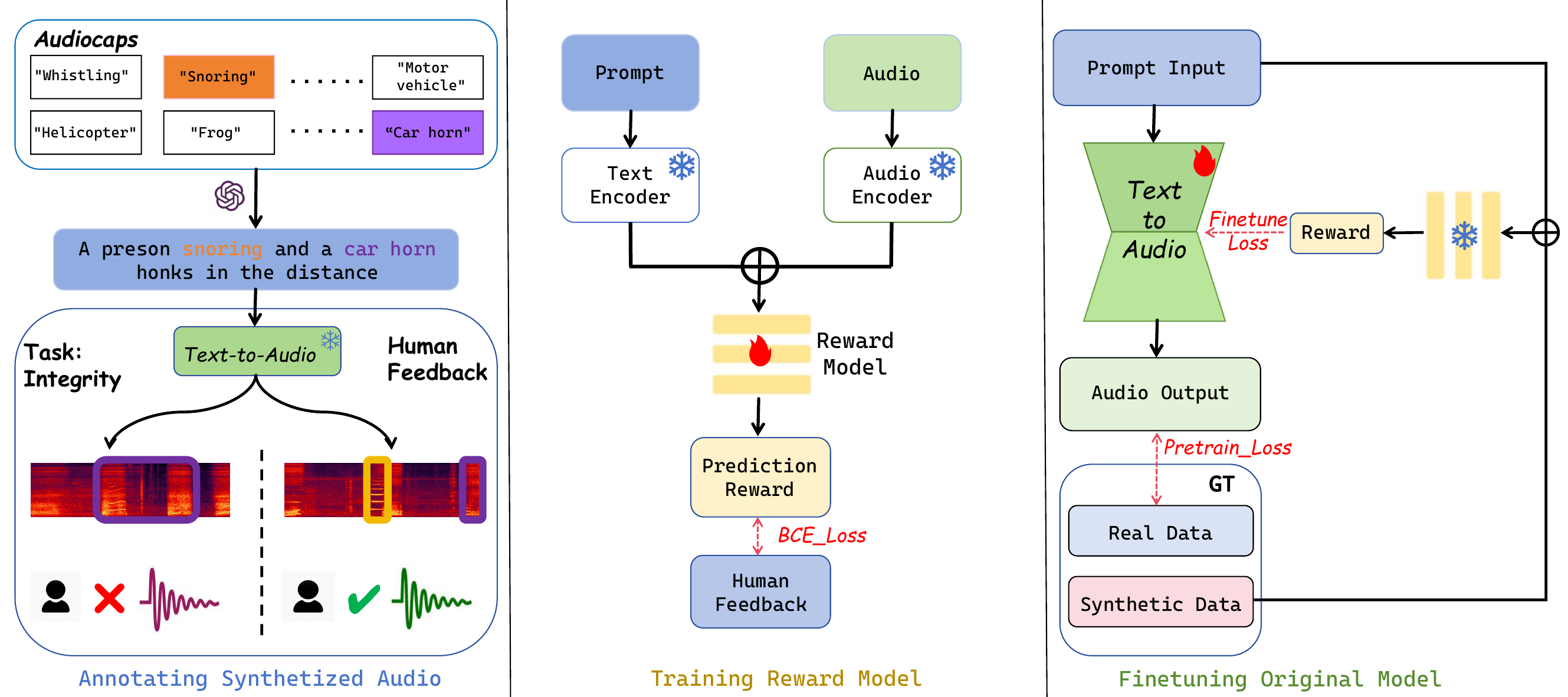}
  \caption{\textbf{The framework of BATON.} BATON integrates three modules: (1) An audio generation unit using LLM-augmented prompts, with human-scored annotations; (2) A reward model trained on synthetic data to emulate human alignment preference; (3) A fine-tuning mechanism that enhance the original generative model using reward model combined human-labeled and pre-training datasets.}
  \label{fig:overview}
    \vspace{-0.4cm}
\end{figure*}

\section{Method}

The framework of BATON, as shown in Fig~\ref{fig:overview}, consists of three components. Firstly, we generated a synthetic dataset of audio-text pairs using selected prompts and then annotated it with task-specific human-derived annotations in Section~\ref{ASA}. Furthermore, we employed the generated data, augmented with human feedback, to train a reward model designed to predict human preferences in audio content, which is illustrated detailedly in Section~\ref{TRM}. Finally, as elaborated in Section~\ref{FOM}, we utilized a reward model reflecting human preferences to fine-tune the text-to-audio model, thereby enhancing its task-specific alignment performance.

\subsection{Dataset Construction}
\label{ASA}

Recent studies~\cite{huang2023makeanaudio,yuan2024retrievalaugmented,xue2024auffusion} have found that text-to-audio models face challenges in producing audio aligning with human preferences, primarily manifest in low integrity, incorrect temporal relationships, and the like. As shown in Figure~\ref{fig:intro-showcase}, the upper example exhibits poor integrity because of a missed audio event and the lower example has an incorrect temporal relationship due to a wrong time order. For simplicity, we here focus on evaluating the integrity and temporal relationships of generated audio to explore the effectiveness of human preference feedback in text-to-audio models. To start, we generate text-audio samples related to these two attributes, and subsequently, human evaluators rate these generated samples.

\begin{table}[t]
\centering
\small
\begin{tcolorbox}[]

Given a label group, $\mathrm{X_{label}}$, the two labels in it will be described into an audio event group: $\{ \texttt{Event}_{1}, \texttt{Event}_{2} \}$. Then, please join the audio event group with conjunction to form an audio caption, defined as "$\texttt{Event}_{1}, \texttt{Conjunction} , \texttt{Event}_{2}$". 
\\
\\
Conjunction list: $\mathrm{X_{conj}}$
\\
\\
Please note that you should randomly select conjunctions from the conjunction list and avoid relying on a single conjunction.
\\
\\
Please try to generate captions that match human language expressions as much as possible.
\end{tcolorbox}
\caption{The generation prompt for the group with 2 labels.}
\label{tab:gpt}
\end{table}

\noindent\textbf{Data Collection.} 
The left part of Figure~\ref{fig:overview} illustrates the process of generating audio data concerning integrity and temporal relationships. We specifically select audio event categories that ranked among the top 200 in occurrence within AudioCaps~\cite{Kim_Kim_Lee_Kim_2019}, constituting our meta labels denoted as $\mathrm{X_{meta}}=\{l_1, ..., l_{200}\}$. Based on that, we randomly select 2 or 3 labels to create a label group, denoted as $\mathrm{X_{label}}$. Conditioned upon the composed $\mathrm{X_{label}}$ and a predefined conjunction list $\mathrm{X_{conj}}$, we instruct the GPT-4, named $\mathrm{M_T}$, with prompt $P$ whereby attaining a complete sentence that matches human language:
\begin{equation}
    \mathbb{D}_{\text{text}} \sim \mathrm{M_T}(\mathrm{P}|\mathrm{X_{label}}, \mathrm{X_{conj}}).
\end{equation}
The prompt $\mathrm{P}$ includes the prefix system message and the suffix prompt, guiding $\mathrm{M_T}$ to produce the desired audio captions. The prompt for the group with 2 labels is detailed in Table~\ref{tab:gpt}. 
The conjunction list $\mathrm{X_{conj}}$ encompasses terms such as \textit{"and"}, \textit{"with"}, \textit{"followed by"}, and so forth (refer to the Appendix for the complete list).
Following the generation process, we manually refine and rephrase certain captions to align them more closely with human oral language. As a result, $\mathbb{D}_{\text{text}}$ encompasses a total of 962 audio captions, with samples presented in Table~\ref{tab:cases}.
Subsequently, we employ the TANGO~\cite{ghosal2023tango}, named $\mathrm{M_A}$, to synthesize 5 different audio for each audio caption:
\begin{equation}
    \mathbb{D}_{\text{data}} \sim \mathrm{M_A}(\mathbb{D}_{\text{text}}).
\end{equation}
Thus, we obtained $4,810$ text-audio pairs in total. 

\begin{table*}[t]
\centering
\begin{tabular}{l|l|l}
\Xhline{0.9pt}
Category                                                                          & \multicolumn{1}{c|}{Labels}                         & \multicolumn{1}{c}{Generated Captions}                                                                                      \\ \Xhline{0.9pt}
\multirow{2}{*}{Integrity}                                                        & "Baby cry, infant cry", "Waves, surf"               & A baby cries while waves crash onto the shore.                                                                              \\
                                                                                  & "Canidae, dogs, wolves", "Rustle"                   & \begin{tabular}[c]{@{}l@{}}Dogs and wolves howl and bark, rustling leaves \\ in the background.\end{tabular}                \\ \hline
\multirow{2}{*}{\begin{tabular}[c]{@{}l@{}}Temproal\\ Relationships\end{tabular}} & "Crying, sobbing", "Toilet flush", "Female singing" & \begin{tabular}[c]{@{}l@{}}Young child crying at first, then a toilet flushes, \\ as a woman's singing begins.\end{tabular} \\
                                                                                  & "Rain on surface", "Helicopter", "Engine starting"  & \begin{tabular}[c]{@{}l@{}}Rain on a surface, followed by a helicopter, \\ and then an engine starts.\end{tabular}          \\ \Xhline{0.9pt}
\end{tabular}
\caption{Examples of generated audio captions.}
\label{tab:cases}
\end{table*}

\noindent\textbf{Human Annotation.}
Upon the collected dataset $\mathbb{D}_{\text{data}}$, we selected 2.7 K samples, denoted as  $\mathcal{D}_\text{human}$, and subjected them to evaluation by human annotators. As presented in Table~\ref{tab:human-dataset}, where $1,431$ samples are related to integrity, and $1,332$ samples are associated with temporal relationships. Annotators employed a binary rating system, assigning either 0 or 1 to each text-audio pair. This is because the nature of integrity or temporal relationship often yields a binary outcome (true or false). Annotators only need to make binary feedback instead of rating a continuous quality score. Admittedly, the more informative human feedback should be adopted for more complex or subjective aspects, such as audio quality and emotional expression. Nevertheless, binary scoring helps to minimize the effects of subjectivity and individual differences for the integrity and temporal relationship tasks, which improves the consistency of the assessment. In practice, we implement a task-specific human preference scoring system, where each audio sample receives an evaluation and is assigned a binary score reflecting its alignment with the given text. A score of \textbf{1} indicates adherence to the desired preference (\textbf{aligned}), whereas a score of \textbf{0} denotes non-conformance (\textbf{not aligned}). We have also set the option \texttt{"}skip\texttt{"} to help the annotators resolve ambiguities of the synthetic audio. As recorded in Table~\ref{tab:human-dataset}, $28.4\%$ text-audio samples in the integrity task are incorrect and $59.4\%$ text-audio samples in the temporal relationships task have wrong order, highlighting a substantial gap between generated audio and human preference.

\begin{table}[t]
\centering
\begin{tabular}{@{}lrrrr@{}}
\toprule
Category     & \multicolumn{1}{l}{Amount of audio} & \multicolumn{3}{c}{Human feedback (\%)} \\ \cmidrule(l){3-5} 
             && \multicolumn{1}{l}{Aligned}  & \multicolumn{1}{l}{Not}         & \multicolumn{1}{l}{Skip} \\ \midrule
Integrity        & \multicolumn{1}{c}{1431}  & \multicolumn{1}{c}{56.1}        & 28.4        & \multicolumn{1}{c}{15.5}  \\
Temporal         & \multicolumn{1}{c}{1332}   & \multicolumn{1}{c}{37.8}        & 59.4        & \multicolumn{1}{c}{2.8}  \\
\midrule
Total            & \multicolumn{1}{c}{2763}      & \multicolumn{1}{c}{47.3}        & 43.4        & \multicolumn{1}{c}{9.3}  \\
\bottomrule
\end{tabular}
\caption{Details of amounts of human feedback dataset.}
\label{tab:human-dataset}
\end{table}



\subsection{Audio Reward Model}
\label{TRM}


As shown in Figure~\ref{fig:overview}, upon the established dataset $\mathcal{D}_\text{human}$, we developed an audio reward model $r_\phi(c,x)$ (parameterized by $\phi$) to predict a scalar reward, which reflects the human preference for the alignment between the provided text $c$ and audio $x$. Such a reward model comprises a text encoder $\mathrm{E}_{\mathrm{C}}(\cdot)$, an audio encoder $\mathrm{E}_{\mathrm{X}}(\cdot)$, and MLP layers. When given the input text $c$ and generated audio $x$, the text encoder and audio encoder extract the respective text embedding $e_c$ and audio embedding $e_x$. Then, $e_c$ and $e_x$ are concatenated along the channel dimension and forwarded to the subsequent MLP layers. The last MLP layer yields a reward that signifies human preference. This process can be summarized as:

\begin{equation}
\begin{aligned}
    &e(c, x) = \mathrm{Cat}(\mathrm{E}_{\mathrm{C}}(c), \mathrm{E}_{\mathrm{X}}(x)), \\
    &r_\phi(c,x) = \mathrm{MLP}(e(c, x)).
\end{aligned}
\end{equation}
In practice, we exploit the audio encoder $\mathrm{E}_{\mathrm{X}}(\cdot)$ and text encoder $\mathrm{E}_{\mathrm{C}}(\cdot)$ from the CLAP model~\cite{wu2023largeclap} that pre-trained on various text-audio samples.
Since the human feedback in $\mathcal{D}_\text{human}$ is binarized, i.e., $y \in \{0, 1\}$, we consider the prediction of human preference as a classification problem. Accordingly, the audio reward model is trained by minimizing the binary cross-entropy loss:

\begin{equation}
\small
    \mathcal{L}(\phi) = \mathop{\mathbb{E}}\limits_{(c, x, y)\sim \mathcal{D}_\text{human}}[-y\log(r_\phi(c,x)) - (1-y)\log(1-r_\phi(c,x))].
\end{equation}
The resulting audio reward model is capable of producing an alignment reward that emulates human preference.


\subsection{Fine-tuning the Text-to-Audio Model with Audio Reward}
\label{FOM}

After obtaining the audio reward model $r_\phi(c,x)$, we combine it with the original conditional audio distribution $p(x|c)$, resulting in a new probability distribution $\tilde p(x|c) = f(r_\phi(c,x))p(x|c)$. Here, $f(\cdot)$ is a monotonically increasing function and we choose $f(\cdot)=exp(\cdot)$ by default. As the original generative text-to-audio model~\cite{ghosal2023tango}, our objective is to update the model $p$ with parameters $\theta$ by maximizing the probability, i.e., minimizing the following loss function:
\begin{equation}
\label{eq:loss_1}
    \mathcal{L}_1(\theta) = \mathbb{E}[-\log \tilde p_\theta(x|c)] = \mathbb{E}[-r_\phi(c,x)\log p_\theta(x|c)].
\end{equation}
Since diffusion models~\cite{ho2020ddpm,rombach2022highldm} are in principle able to model conditional distributions with a conditional denoising autoencoder $\epsilon_\theta(x_t,t,c)$, the Eq.~\eqref{eq:loss_1} can be simplified as:
\begin{equation}
\label{eq:loss_6}
    \mathcal{L}_2(\theta) = \mathbb{E}_{t,(x_t,c)\sim \mathcal{D}_\text{data}}[r_\phi(c,x_t)\|\epsilon-\epsilon_\theta(x_t,c,t)\|^2],
\end{equation}
where $x_t$ is the noise version of the input $x$.
Compared to the original denoising loss, the reward term behaves as a modulating factor: when an input sample aligns with human preference, the modulating factor is large, thus the text-to-audio model pays more attention to learning this good sample, and vice versa.
In addition, since our constructed dataset is relatively small compared to the pre-trained dataset, we introduce the denoising loss from the original text-to-audio model~\cite{ghosal2023tango}.
Consequently, the final loss is:
\begin{align}
    \mathcal{L}(\theta) = &\mathbb{E}_{t,(x_t,c)\sim \mathcal{D}_\text{data}}[r_\phi(c,x_t)\|\epsilon-\epsilon_\theta(x_t,c,t)\|^2] \notag \\
&+ \beta \mathbb{E}_{t,(x_t,c)\sim \mathcal{D}_\text{pretrain}}[\|\epsilon-\epsilon_\theta(x_t,c,t)\|^2],
\end{align}
in which $\mathcal{D}_\text{pretrain}$ refers to the sampled data from the pre-trained dataset~\cite{Kim_Kim_Lee_Kim_2019} and $\beta$ is a hyper-parameter.
This regularized loss prevents the model from overfitting to our self-build dataset, ensuring it does not deviate heavily from the original model. 

\section{Experiments}
\begin{table*}[htbp]
\centering
\begin{tabularx}{\textwidth}{l|l|XXXXX|ll}
\toprule
     Task &    Model &     FD$\downarrow$ & FAD$\downarrow$ &  IS$\uparrow$ & KL$\downarrow$ & S$_{CLAP}$$\uparrow$ & MOS-Q$\uparrow$& MOS-F$\uparrow$ \\
\midrule
\multirow{4}{*}{\small{Integrity}} &  \small{TANGO$^{\dag}$\cite{ghosal2023tango}} &   \textbf{36.79} &  \textbf{2.81} &    4.78 &  \textbf{1.03} &  33.0$\%$ &   4.05±\small{0.07} &   3.70±\small{0.07} \\
          &  \small{AudioLDM-L\cite{liu2023audioldm}} &   64.52 &  7.28 &  3.77 &  2.64 &   18.5\% &   3.70±\small{0.08} &   2.95±\small{0.08} \\
          & \small{AudioLDM2-L\cite{liu2023audioldm2}} &  44.01 &  3.13 &  4.84 &   1.44 &  28.8\% &   3.33±\small{0.08} &   3.34±\small{0.08} \\
          &   BATON (ours) &  38.54 &   3.02 &   \textbf{5.07} &  1.11 &  \textbf{35.3$\%$} &   \textbf{4.55±\small{0.05}} &   \textbf{4.42±\small{0.05}} \\
          \midrule
\multirow{4}{*}{\small{Temporal}} &  \small{TANGO$^{\dag}$\cite{ghosal2023tango}} &  36.86 &  4.02 &  4.46 &  1.16 &   35.9\% &4.15±\small{0.06} &3.83±\small{0.06}\\
          &  \small{AudioLDM-L\cite{liu2023audioldm}} &  65.83 &  7.21 &  3.89 &  2.74 &  19.5\% &3.39±\small{0.07}&2.79±\small{0.08}\\
          & \small{AudioLDM2-L\cite{liu2023audioldm2}} &  45.34 &  \textbf{3.05} &  4.69 &  1.50 &  31.9\% & 3.19±\small{0.07} & 3.51±\small{0.08}\\
          & BATON (ours) &   \textbf{36.19} &  4.31 &  \textbf{4.88} &  \textbf{1.13} &  \textbf{41.9$\%$} & \textbf{4.68±\small{0.05}} &\textbf{4.41±\small{0.05}} \\
\bottomrule
\end{tabularx}
\caption{\textbf{Comparison of different text-to-audio models.} TANGO$^{\dag}$ is the baseline model which is ultilized by BATON to fine-tune with human preference feedback.}
\label{main result}
\end{table*}

\subsection{Implementation Details}
Employing TANGO (Full-FT-Audiocaps)~\cite{ghosal2023tango} as our baseline, we integrated a human feedback-driven fine-tuning framework. From the Audiocaps dataset, top occurrences 200 labels were chosen, and GPT-4~\cite{openai2023gpt4} was utilized to expand these into prompts for text-to-audio synthesis, resulting in a dataset of 962 audio captions and their 4810 resultant audio outputs of which default length is set to 10 seconds. Human annotators then binary score the text-audio pairs, assessing their alignment with preference criteria. We trained the audio reward model on synthetic dataset over 50 epochs, with a batch size 64 and learning rate 0.01 using the Adam~\cite{Kingma_Ba_2014}. 
During the fine-tuning of the original model, we assigned a weight parameter $\beta$ of 0.5 to the pretrain loss. The ratio of human-labeled data to reward model-scored data was maintained at 1:1 for both sub-tasks. $\mathcal{D}_\text{data}$ and $\mathcal{D}_\text{pretrain}$ involve $4.8$ K and $2.5$K samples, respectively. The fine-tuning was conducted over 10 epochs, with settings for the learning rate $1 \times 10^{-5}$, batch size 6, and default optimizer AdamW~\cite{Loshchilov_Hutter_2017}. 
\subsection{Main Result}

\textbf{Evaluation method.} For evaluation, we filtered the Audiocaps test set using words such as \textit{and, as, then, while, before, after, followed} to obtain a multi-labeled test set. Subsequently, we curated subsets comprising 148 two-label and 165 three-label prompts from the filtered data. These two subsets are used to evaluate audio integrity and temporal consistency respectively. Consistent with prior research~\cite{huang2023makeanaudio,liu2023audioldm,liu2023audioldm2,xue2024auffusion}, our objective evaluation of audio quality and fidelity employs metrics such as Fréchet Distance (FD), Fréchet Audio Distance (FAD)~\cite{Kilgour_Zuluaga_Roblek_Sharifi_2019}, Kullback-Leibler divergence (KL)~\cite{yang2023diffsound}. We adapt Inception Score (IS) to assess the quality and variety of samples. Additionally, we utilized the Cross-Modal Language-Audio Perceptual (CLAP) metric S$_{CLAP}$ to objectively assess the alignment between audio and text. For subjective analysis, 40 participants were recruited to rate the perceived quality and text alignment of the audio separately on a 5-point scale, with 5 being the highest possible score. The Mean Opinion Score (MOS) method, utilizing a $96\%$ confidence interval, is applied to assess both the audio quality and the faithfulness of text-to-audio alignment. These assessments are quantified as MOS-Q and MOS-F, respectively.

\noindent\textbf{Quantitative experiment.} BATON showcases competitive performance across multiple metrics in text-to-audio tasks, as detailed in Table~\ref{main result}. In evaluating the similarity between generated and real samples, the FAD metric exhibits a slight reduction compared to the original model in both tasks. Considering that FD relies on PANNs~\cite{Kong_Cao_Iqbal_Wang_Wang_Plumbley_2020} classifiers and the labels in our fine-tuned dataset constitute a subset of the PANNs sound class, using FD as the primary metric for assessing audio quality appears to be more reliable. On the temporal order task, the fine-tuned model surpasses the original model in terms of FD, IS and KL, indicative of its robustness in maintaining semantic cohesion while minimizing divergence from the target distribution. Notably, the BATON model surpasses other models in both subtasks for the IS metric, suggesting its excellent capability in audio variety. The superior performance of BATON in $S_{\text{CLAP}}$, particularly a 2.3\% improvement over the baseline TANGO model in the integrity task and a significant 6.0\% enhancement in the temporal task, underscores its effectiveness in producing audio that aligns closely with human preferences. This enhancement can be attributed to our fine-tuning approach that leverages human feedback, allowing BATON to better understand and distinguish the nuances of audio content that are valued by human listeners. 

When subjectively evaluated on the integrity task, BATON demonstrates a superior MOS-Q of 4.55±0.05, outperforming TANGO and AudioLDM-L(4.05±0.07 and 3.70±0.08 respectively). Concurrently, BATON's MOS-F is 4.42±0.05, indicating a more faithful text-to-audio alignment compared to TANGO's 3.70±0.07. Significantly, BATON continues to demonstrate superiority in subjective metrics for the temporal task, with its MOS-F surpassing the original model by 0.58 points. This observation indicates its adept handling of temporal information. These findings affirm that BATON not only generates high-quality audio that maintains or even exceeds the original model, but also sustains an audio reconstruction of the textual details.
\begin{figure*}[t]
  \centering
  \includegraphics[width=\linewidth]
  {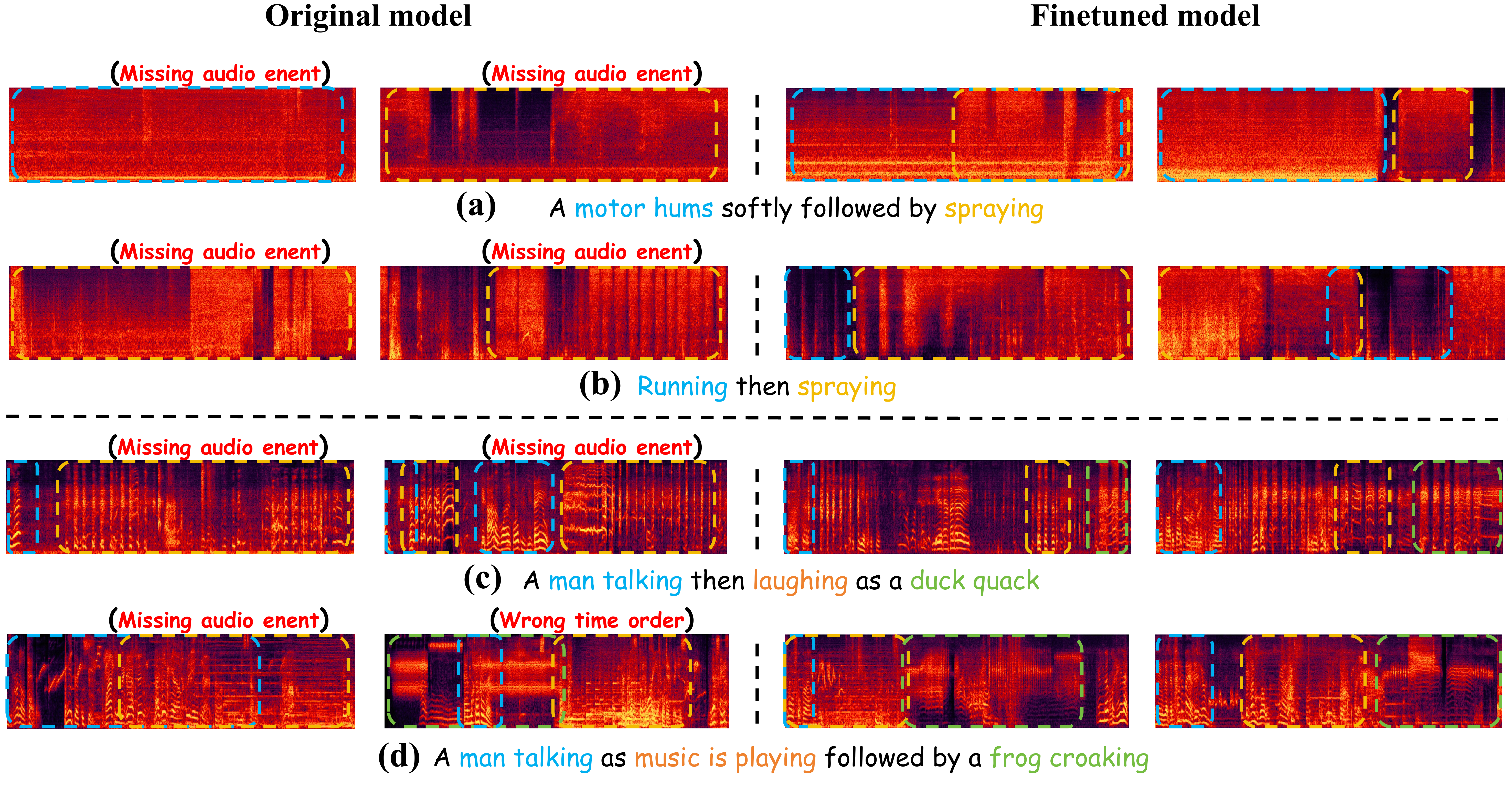}
  \caption{Generated samples comparison of TANGO (original model) and BATON (finetuned model). The left two samples in the display are from the original model, while the right two are from the post-finetuned model. Comparisons (a) and (b) show that the finetuned model produces complete audio events, unlike the original model which omits certain audio event. In comparisons (c) and (d), the original model generates audio with a confused sequence, whereas the finetuned model adheres to the sequence of prompt.}
  \label{fig:exp}
\end{figure*}

\noindent\textbf{Qualitative experiment.} Figure~\ref{fig:exp} compares the original and fine-tuned models in terms of generation faithfulness for two-label integrity and three-label temporal samples. In our qualitative analysis, discernible discrepancies are noted in the audio generation by the original model, especially for two-label samples. For instance, in response to the prompt "\textit{A \textbf{motor hums} softly followed by \textbf{spraying}}", the generated audio often included only one of the elements, such as "\textit{motor hums}" or "\textit{spraying}", thus omitting specific audio event. Similarly, for three-label samples, the original model misrepresented the sequence of events. Example prompt "\textit{A \textbf{man talking} as \textbf{music is playing} followed by a \textbf{frog croaking}}", where the generated audio incorrectly prioritized the "\textit{frog croaking}" event over the initial "\textit{man talking}". In contrast, BATON demonstrates enhanced performance, achieving more faithful generation on diverse audio events and temporal order shown in prompts.

\subsection{Ablation Study}
Considering the heightened intricacy posed by the three-label temporal task, which involves increased complexity in handling more audio events and challenges in temporal alignment, our ablation examination focuses on evaluating the model specifically on the test set of the second subtask.

\noindent\textbf{Human feedback.} As detailed in Table~\ref{ablation of RLHF}, to dissect whether the enhancement in model performance is attributed to human feedback or increased data volume, we conducted a comparative analysis of the impact of data with and without human annotations on the model's performance. In instances without human annotations, all data were used to fine-tune the model based on pretrain loss only. In the case with human annotations, we opted for BATON, for comparative evaluation. The baseline model, without fine-tuning, achieved CLAP score of 35.9\%. Incorporating PD yielded slight gains in KL and CLAP accuracy but FD and FAD decreased, while continually integrating synthetic dataset which including human-annotated data and reward model labeled data to pretrain data progressively improved CLAP score to 36.2\% and 37.5\%. Fine-tuning with human feedback achieves the highest CLAP accuracy at 41.9\%, demonstrating this uptick is primarily attributed to the feedback of human preference instead of the boost of training data volume.

\begin{table}[h]
\centering
\begin{tabularx}{0.48\textwidth}{XXXX|XXXX|l}
\toprule
\small{PD} & \small{HD$^{\ddag}$} & \small{RD$^{\ddag}$} & \small{HF} & \small{FD$\downarrow$} & \small{FAD$\downarrow$} & \small{IS$\uparrow$} & \small{KL$\downarrow$} & \small{S$_{CLAP}$$\uparrow$}\\
\midrule
           &            &            && \small{36.86}  & \small{\textbf{4.02}} & \small{4.46} & \small{1.17} &35.9\%\\
\checkmark &            &            && \small{37.06} & \small{4.08} & \small{4.45} & \small{1.11} &36.1\%\\
\checkmark & \checkmark &            && \small{37.78}  & \small{4.49} & \small{4.70} & \small{\textbf{1.09}} &36.2\%\\
\checkmark & \checkmark & \checkmark && \small{37.23}  & \small{4.09} & \small{4.80} & \small{1.15} &37.5\%\\
           &            &             &\checkmark & \small{\textbf{36.19}}  & \small{4.31} & \small{\textbf{4.88}} & \small{1.13} &\textbf{41.9\%}\\
\bottomrule
\end{tabularx}
\caption{Ablation study for efficacy of human feedback on the test set. 'PD', 'HD$^{\ddag}$' and 'RD$^{\ddag}$' separately represents for fine-tuning with pretrain data, human-annotated data and reward model labeled data using pretrain loss and no preference feedback value.}
\label{ablation of RLHF}
\vspace{-2mm}
\end{table}

\noindent\textbf{Preference data.} The ablation study detailed in Table~\ref{Ablation of preference} investigates the influence of different preference data sources on BATON. Initially, employing only human preference data provides a CLAP score of 38.8\%, indicating a moderate enhancement over the baseline. The exclusive use of reward model annotating data achieves an improved CLAP accuracy of 38.8\%, suggesting that reward model data alone is beneficial as well. The most substantial improvements were observed when both human annotations and reward model annotations are utilized together, optimizing the performance with an FD score of 36.19 and a peak CLAP score of 41.9\%. This configuration demonstrates the synergistic effect of combining human annotation with reward model assessment, significantly improving the generation with human preference.

\begin{table}[h]
\centering
\begin{tabularx}{0.45\textwidth}{cc|XXXX|X}
\toprule
\small{HA} & \small{RA} & \small{FD$\downarrow$} & \small{FAD$\downarrow$} & \small{IS$\uparrow$} & \small{KL$\downarrow$} & \small{S$_{CLAP}$$\uparrow$}\\
\midrule
           &            & 36.86  & \textbf{4.02} & 4.46 & 1.17 &35.9\%\\
\checkmark &            & 37.95 & 4.80 & 4.87 & \textbf{1.07} &38.8\%\\
           & \checkmark & 37.50  & 4.52 & \textbf{4.88} & 1.14 &38.8\%\\
\checkmark & \checkmark & \textbf{36.19}  & 4.31 & \textbf{4.88} & 1.13 &\textbf{41.9\%}\\
\bottomrule
\end{tabularx}
\caption{Ablation study for preference data source on the test set. 'HA' and 'RA' denote utilizing human-annotated data and reward model labeled data.}
\label{Ablation of preference}
\vspace{-1mm}
\end{table}

\noindent\textbf{Pretrain loss.} The ablation study, outlined in Table~\ref{Ablation of pretrain_loss} evaluates the efficacy of pretrain loss in fine-tuning. Incorporating pretrain loss improved the FD and FAD score to 36.19 and 4.31 while IS and KL score slightly decrease, and CLAP score rises to 41.9\%, underscoring its significance in enhancing alignment and preventing overfitting on a relatively small amount of preference data.

\begin{table}[h]
\centering
\begin{tabularx}{0.45\textwidth}{c|XXXX|X}
\toprule
\small{PL} & \small{FD$\downarrow$} & \small{FAD$\downarrow$} & \small{IS$\uparrow$} & \small{KL$\downarrow$} & \small{S$_{CLAP}$$\uparrow$}\\
\midrule
            w/o& 36.93  & 4.48 & \textbf{4.99} & \textbf{1.07} &39.7\%\\
            w& \textbf{36.19} & \textbf{4.31} & 4.88 & 1.13 &\textbf{41.9\%}\\
\bottomrule
\end{tabularx}
\caption{Ablation study for pretrain loss on the test set. 'PL' represents for fine-tuning with pretrain loss.}
\label{Ablation of pretrain_loss}
\vspace{-1mm}
\end{table}

\noindent\textbf{Reward model.} As shown in Table~\ref{Ablation of encoder&loss}, comparison of encoders and training losses for the audio reward model reveals that the CLAP text encoder, coupled with BCE loss, significantly outperforms the Flan-T5~\cite{shen2023flant5} encoder in test set performance, yielding superior alignment outcomes. This efficacy likely stems from ability of CLAP, through large-scale pretraining based on contrastive learning, to extract more aligned text feature and audio feature, and from aptness of BCE loss for binary classification (0 and 1) tasks within reward model training.

\begin{table}[h]
\centering
\begin{tabularx}{0.36\textwidth}{XX|XX|l}
\toprule
\small{CLAP} & \small{T5} &\small{MSE} & \small{BCE} & S$_{CLAP}$$\uparrow$\\
\midrule
           & \checkmark&\checkmark &      & \multicolumn{1}{c}{40.9\%}\\
           & \checkmark &       &\checkmark & \multicolumn{1}{c}{40.9\%}\\
\checkmark &            &\checkmark&     &\multicolumn{1}{c}{41.0\%}\\
\checkmark &            &       &\checkmark      &\multicolumn{1}{c}{\textbf{41.9\%}}\\
\bottomrule
\end{tabularx}
\caption{Ablation study for feature extractors and loss adapted in reward model on the test set.}
\label{Ablation of encoder&loss}
\vspace{-1.5mm}
\end{table}

\begin{figure}[h]
  \centering
  \begin{subfigure}{0.23\textwidth}
    \centering
    \includegraphics[width=\linewidth]{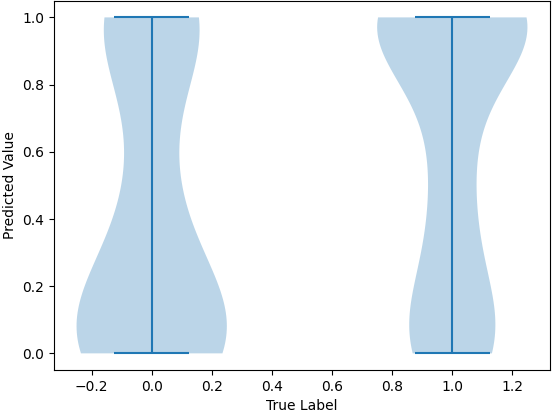}
    \caption{Integrity test set}
    \label{subfig1}
  \end{subfigure}\hfill
  \begin{subfigure}{0.23\textwidth}
    \centering
    \includegraphics[width=\linewidth]{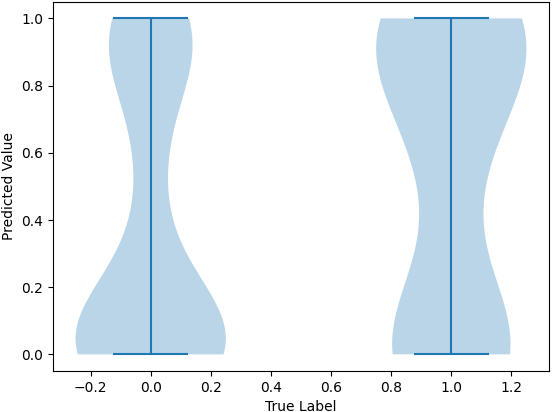}
    \caption{Temporal order test set}
    \label{subfig2}
  \end{subfigure}
  \caption{Prediction distribution of audio reward models.}
  \label{fig:combined}
  \vspace{-2mm}
\end{figure}

To assess the accuracy of reward model in predicting human preferences, we applied it to two subtasks test sets, each comprising 120 text-audio pairs, and annotated by two separate individuals. Figure~\ref{fig:combined} presents the findings. The predictions prominently feature two high-frequency peaks at 0 ("Not Aligned") and 1 ("Aligned"), suggesting a strong alignment of the model with human ratings. The concentration of predictions at these extremes, with few near the mid-point of 0.5, indicates the model's effectiveness in differentiating text-audio pair quality, particularly in tasks of integrity and temporal relationships.

\section{Conclusion and Discussion}
\textbf{Conclusion.} 
In this paper, we propose a novel framework (BATON), using human feedback to enhance the alignment between text prompts and generated audio in TTA models. Our BATON involves three steps: collecting human feedback for the constructed text-audio pairs, training an audio reward model on human-annotated data, and fine-tuning an off-the-shelf TTA model using the audio reward model. Extensive experiments manifest that BATON effectively leverages human feedback to mitigate biases and improve alignment in TTA models, making a valuable contribution to the ongoing advancements in audio synthesis from textual inputs. We anticipate that BATON can pave the way for aligning TTA models through human feedback.


\noindent\textbf{Discussion.} As a plug-and-play fine-tuning framework for various TTA models, BATON has some limitations and areas for future exploration. Firstly, the data-driven characteristic of BATON means the performance of alignment depends on the quality and quantity of human feedback. Secondly, the reliance of our approach on a two-stage training process, rather than using reinforcement learning strategies for online fine-tuning, restricting BATON to offline updates. 
Future work should further explore reinforcement learning with human feedback in text-audio alignment.


\bibliographystyle{named}
\bibliography{ijcai24}

\clearpage
\appendix
\begin{center}{\bf \Large Appendix}\end{center}\vspace{-2mm}

\section{Project Page}
Project page is available at \url{https://baton2024.github.io}.

\section{Details of Fine-tuning the Text-to-Audio Model with Reword Model}



In this section, we provide more details about fine-tuning diffusion models with the reward model. For the new probability distribution $\tilde p(x|c) = f(\cdot)p(x|c) = f(r_\phi(c,x))p(x|c)$ established in Section 3.3, which integrates the reward model and the original probability distribution, we define the function $f(\cdot)$ according to the control-as-inference graphical theory \cite{2018Reinforcement}. If we only consider a one-step MDP scenario, we can define a random variable $\mathcal{O}$ denoting the optimality of the audio $x$ and $p(\mathcal{O}=1|c,x)\propto \text{exp}(r_\phi(c,x))$. In order to increase the probability of generating optimal audio, we choose $f(r_\phi(c,x))=\text{exp}(r_\phi(c,x))$. Following the paradigm of diffusion training and guided by the principles of maximum likelihood estimation, our objective is to maximize the probability of $\tilde p_\theta(x|c)$, equivalent to minimizing Eq. \eqref{eq:loss_1}.

If the reward model is a nonnegative function, the evidence lower bound (ELBO) of this distribution can be written as:
\begin{align*}
& \scalebox{1.1}{$-r_\phi(c,x)\log p_{\theta}(x|c)$} \\
&\leq -r_\phi(c,x)\log p_{\theta}(x|c) \\
&\quad + r_\phi(c,x) \scalebox{0.96}{$D_{\mathrm{KL}}\left(q\left(x_{1: T} \mid x,c\right) \| p_{\theta}\left(x_{1: T} \mid x,c\right)\right)$} \\
&= -r_\phi(c,x)\log p_{\theta}(x|c) \\
&\quad + r_\phi(c,x)\scalebox{1.02}{$\mathbb{E}_{x_{1: T} \sim q\left(x_{1: T} \mid x,c\right)}\left[\log \frac{q\left(x_{1: T} \mid x,c\right)}{p_{\theta}\left(x_{0: T}|c\right) / p_{\theta}\left(x|c\right)}\right]$} \\
&= -r_\phi(c,x)\log p_{\theta}(x|c) \\
&\quad + \scalebox{0.96}{$r_\phi(c,x)\mathbb{E}{q}\left[\log \frac{q\left(x_{1: T} \mid x,c\right)}{p_{\theta}\left(x_{0: T}|c\right)} + \log p_{\theta}(x|c)\right]$} \\
&= r_\phi(c,x)\scalebox{0.96}{$\mathbb{E}{q}\left[\log \frac{q\left(x_{1: T} \mid x,c\right)}{p_{\theta}\left(x_{0: T}|c\right)}\right]$}.
\end{align*}

In fact, the label we train is either 0 or 1, so we constrain the predicted values to be between 0 and 1, ensuring the nonnegative output of the reward model $r_\phi(c,x)$. The expectation $\mathbb{E}_{q}\left[\log \frac{q\left(x_{1: T} \mid x,c\right)}{p_{\theta}\left(x_{0: T}|c\right)}\right]$ is the same as the classifier-guided diffusion model's expectation \cite{prafulla2021diffusion}. By using the format of $L_\text{simple}$ in the work, the loss can be written as Eq. \eqref{eq:loss_6}.


\section{Generate Captions}

\subsection{Label Set}

According to the metadata "label" within the training set of AudioCaps~\cite{Kim_Kim_Lee_Kim_2019}, the frequency of each label is tallied, and the top 200 labels were selected for making label set. The top 50 are shown in the Table~\ref{tab:example}.

\begin{table*}[htbp]
    \centering
    \begin{tabularx}{\textwidth}{Xp{10cm}}
        \toprule
        \multicolumn{1}{c}{\small{\textbf{Label Set}}} \\
        \midrule
        \small{"Speech", "Vehicle", "Animal", "Car", "Domestic animals, pets", "Narration, monologue", "Dog", "Bird", "Inside, small room", "Train", "Rail transport", "Male speech, man speaking", "Train horn", "Railroad car, train wagon", "Female speech, woman speaking", "Bow-wow", "Outside, urban or manmade", "Engine", "Motor vehicle (road)", "Boat, Water vehicle", "Music", "Outside, rural or natural", "Stream", "Truck", "Car passing by", "Accelerating, revving, vroom", "Horse", "Motorboat, speedboat", "Whimper (dog)", "Door", "Water", "Idling", "Wind", "Gurgling", "Helicopter", "Clip-clop", "Rain", "Hiss", "Clickety-clack", "Bus", "Wood", "Snoring", "Race car, auto racing", "Aircraft", "Tick-tock", "Motorcycle", "Emergency vehicle", "Pigeon, dove", "Spray", "Duck"} \\
        \bottomrule
    \end{tabularx}
    \caption{The top 50 most frequent labels in AudioCaps Training set}
    \label{tab:example}
\end{table*}

\subsection{Prompt Generation}

We randomly select 2 or 3 labels from the label set to form a label group $\mathrm{X_{label}}$, then encourage GPT-4~\cite{openai2023gpt4} to generate label-based audio captions. The designed instruction for the two tasks shown in Table~\ref{GPT Instruction1} and~\ref{GPT Instruction2}, respectively. 

\begin{table}[H]
\centering
\small
\begin{tcolorbox}[]

Given a label group, $\mathrm{X_{label}} = \{ \texttt{label}_{1}, \texttt{label}_{2} \}$, the two labels in it will be described into an audio event group: $\{ \texttt{Event}_{1}, \texttt{Event}_{2} \}$. Then, please join the audio event group with conjunction from conjunction list to form an audio caption. The generated answer should follow the format "$\texttt{Caption}:\texttt{"Event}_{1}, \texttt{Conjunction}, \texttt{Event}_{2}\texttt{"}$, $\texttt{Label}:\{ \texttt{label}_{1}, \texttt{label}_{2} \}$". 
\\
\\
Conjunction list: [$,$], [$\texttt{and}$], [$\texttt{while}$], [$\texttt{with}$], [$\texttt{as}$], [$\texttt{followed by}$], [$\texttt{then}$], [$\texttt{and then}$], [$\texttt{before}$]
\\
\\
Please note that you should randomly select conjunctions from the conjunction list and avoid relying on a single conjunction.
\\
\\
Please try to generate captions that match human language expressions as much as possible.
\end{tcolorbox}
\vspace{-0.3cm}
\caption{GPT instruction for integrity task.}
\label{GPT Instruction1}
\vspace{-0.6cm}
\end{table}

\begin{table}[H]
\centering
\small
\begin{tcolorbox}[]
Given a label group, $\mathrm{X_{\text{label}}} = \{ \texttt{label}_{1}, \allowbreak \texttt{label}_{2}, \allowbreak \texttt{label}_{3} \}$, the three labels in it will be described into an audio event group: $\{ \texttt{Event}_{1}, \texttt{Event}_{2},\texttt{Event}_{3} \}$. Then, please join the audio event group with conjunction from conjunction list to form an audio caption. The generated answer should follow the format "$\texttt{Caption}:\texttt{"Event}_{1}, \texttt{Conjunction}_{1}, \texttt{Event}_{2}, \texttt{Conjunction}_{2}, \allowbreak \texttt{"Event}_{3}\texttt{"}$, $\texttt{Label}:\{ \texttt{label}_{1}, \texttt{label}_{2}, \texttt{label}_{3} \}$". 
\\
\\
Conjunction list: [\textless $\texttt{followed by} \textgreater, \textless \texttt{followed by} \textgreater$], [\textless $\texttt{followed by} \textgreater, \textless \texttt{and then} \textgreater$], [\textless $\texttt{followed}  \newline \texttt{by} \textgreater, \textless \texttt{then} \textgreater$], [\textless $\texttt{then} \textgreater, \textless \texttt{followed by} \textgreater$], [\textless $\texttt{be}  \newline \texttt{fore} \textgreater, \textless \texttt{followed by} \textgreater$], [\textless $\texttt{and then} \textgreater, \textless \texttt{and} \textgreater$ \newline], [\textless $\texttt{then} \textgreater, \textless \texttt{and} \textgreater$], [\textless $\texttt{followed by} \textgreater, \textless \texttt{and} \textgreater$], [\textless $\texttt{followed by} \textgreater, \textless \texttt{as} \textgreater$], [\textless $\texttt{before} \textgreater, \textless \texttt{as} \textgreater$], [\textless $\texttt{with} \textgreater, \textless \texttt{and then} \textgreater$]
\\
\\
Please note that you should randomly select conjunctions from the conjunction list and avoid relying on a single conjunction.
\\
\\
Please try to generate captions that match human language expressions as much as possible.

\end{tcolorbox}
\vspace{-0.3cm}
\caption{GPT instruction for temporal task.}
\label{GPT Instruction2}
\vspace{-0.4cm}
\end{table}

\section{Human Annotation and Evaluation}

\subsection{Human Annotation}

The two-label integrity and three-label temporal tasks were annotated separately by different groups of annotators. Annotators were provided with five concurrently generated audio for each prompt. The annotation rules are as follows: 

\begin{itemize}
    \item \textbf{Integrity}: Please aurally recognize whether both audio events of the prompt are present, regardless of whether the temporal order or frequency of audio events corresponds to the prompt. Score \textbf{0} if one or more audio events are absent; score \textbf{1} when both audio events are present. In cases where it is challenging to discern, assign \textbf{Uncertain}.
    \item \textbf{Temporal}: Not only do you need to determine if all three audio events in the prompt have occurred, but you also need to identify whether the temporal order of the generated audio corresponds to the prompt. Score \textbf{0} if one or more audio events are absent or if the temporal order are inconsistent with the prompt; score \textbf{1} when all three audio events are present, and their temporal order are consistent with the prompt. In cases where it is challenging to discern, assign \textbf{Uncertain}.
\end{itemize}

\begin{figure*}[htbp]
    \centering
    \includegraphics[width=0.9\textwidth]{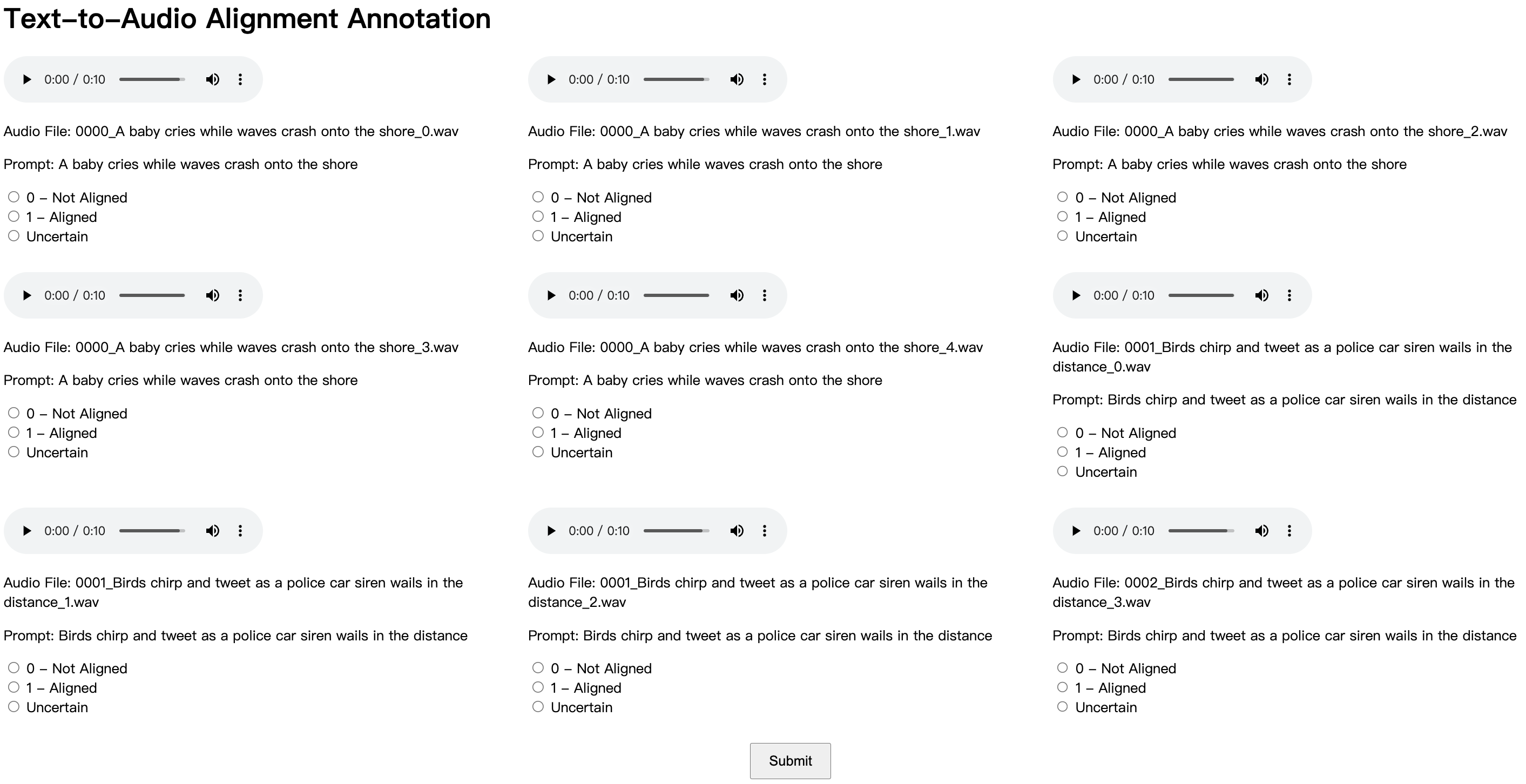}
    \captionsetup{justification=centering}
    \caption{Screenshot of annotaion system.} 
    \label{Annotaion system}
\end{figure*}

Our annotation page screenshot is displayed in Figure \ref{Annotaion system}.

\subsection{Human Evaluation}

To evaluate the impact of human feedback-based fine-tuning on the alignment and quality of the generated audio, we engaged 40 participants, divided into two groups, to perform Mean Opinion Score (MOS) tests. One group assessed a two-label integrity task sample, while the other group evaluated a three-label temporal task sample. Both groups were tasked with assessing alignment and quality metrics.

\begin{figure}[H]
    \centering
    \includegraphics[width=0.5\linewidth]{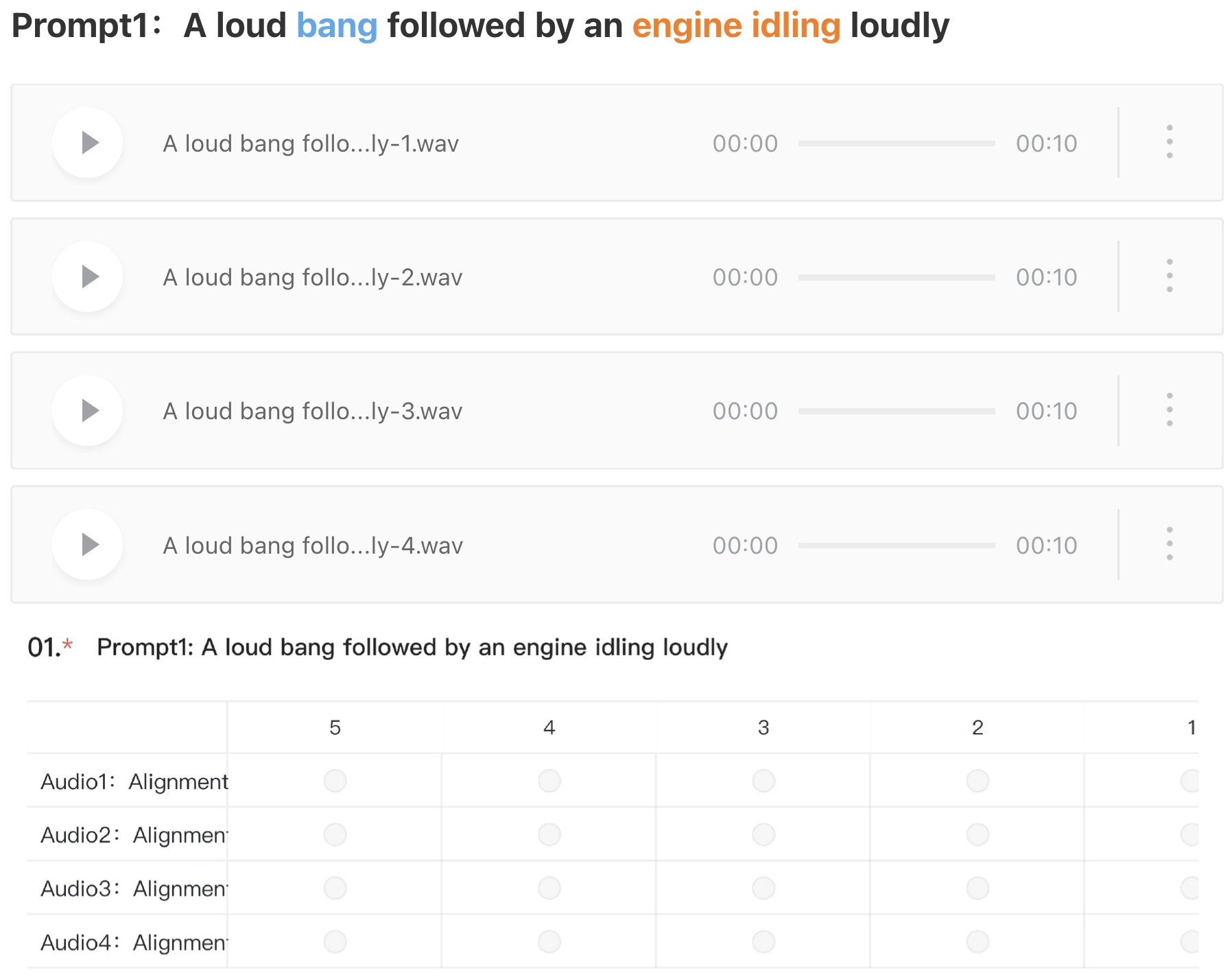}
    \captionsetup{justification=centering}
    \caption{Screenshot of audio alignment evaluation system.}
    \label{fig:evaluation system1}
    \vspace{-0.3cm}
\end{figure}

\begin{figure}[H]
    \centering
    \includegraphics[width=0.5\linewidth]{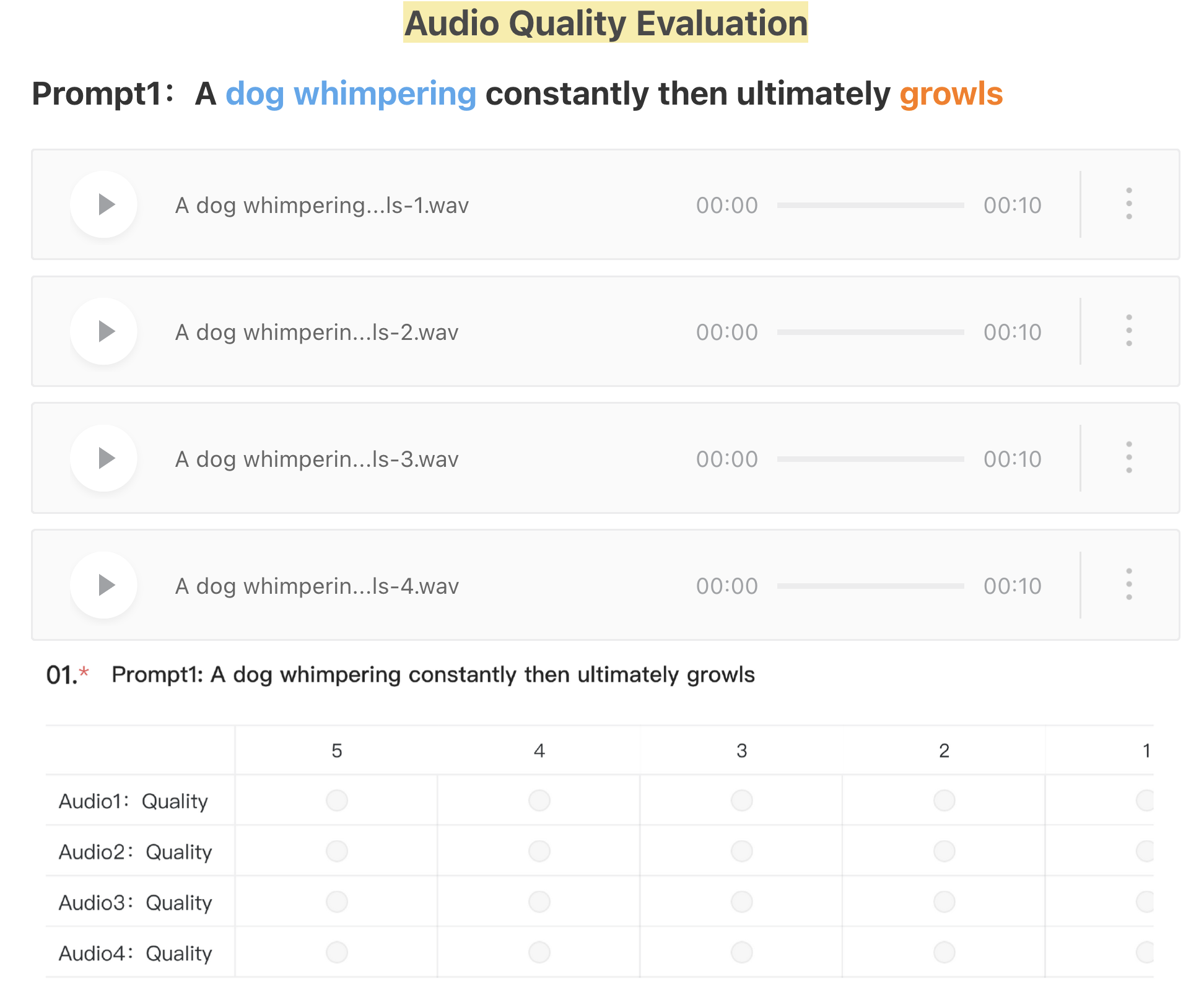}
    \captionsetup{justification=centering}
    \caption{Screenshot of audio quality evaluation system.}
    \label{fig:evaluation system2}
    \vspace{-0.3cm}
\end{figure}

The evaluation of two metrics was conducted a Likert scale, spanning a range from one to five, and Table~\ref{tab:score tab} illustrates the rating scales in the respective tasks. The test questionnaire for audio alignment and quality is presented in the Figure~\ref{fig:evaluation system1} and Figure~\ref{fig:evaluation system2}, 

\begin{table*}[ht]
    \centering
    \begin{tabularx}{\linewidth}{>{\centering}p{1.5cm}|>{\centering}p{0.8cm}|>{\raggedright}p{6.0cm}|>{\raggedright\arraybackslash}X}
        \toprule
        \small{Task} & \small{Score} & \small{Alignment} & \small{Quality} \\
        \midrule
        \small{Integrity} & 5 & Both sound events are generated well & Clear and natural audio \\
                   & 4 & One audio event seems to be missing & Relatively natural audio, overall satisfactory \\
                   & 3 & One sound event is clearly missing & Audio exhibits obvious imperfections, but acceptable \\
                   & 2 & One sound event is clearly missing and \\ 
                   the other may be missing & Poorer audio quality and auditory experience \\
                   & 1 & Two sound events are clearly missing & Extremely poor audio, almost unacceptable \\
        \midrule
        \small{Temporal}  & 5 & Three sound events are generated well and \\
        in perfect temporal order & Clear and natural audio \\
                   & 4 & Three sound events are generated well but \\
                   out of temporal order & Relatively natural audio, overall satisfactory \\
                   & 3 & One sound event is missing & Audio exhibits obvious imperfections, but acceptable \\
                   & 2 & Two sound events are missing & Poorer audio quality and auditory experience \\
                   & 1 & Three sound events are missing & Extremely poor audio, almost unacceptable \\
        \bottomrule
    \end{tabularx}
    \caption{Rating scales for audio alignment and quality evaluation.}
    \label{tab:score tab}
\end{table*}

\section{Additional Results}

\subsection{Number of iterations for fine-tuning}

\begin{figure}[H]
    \centering
    \includegraphics[width=0.99\linewidth]{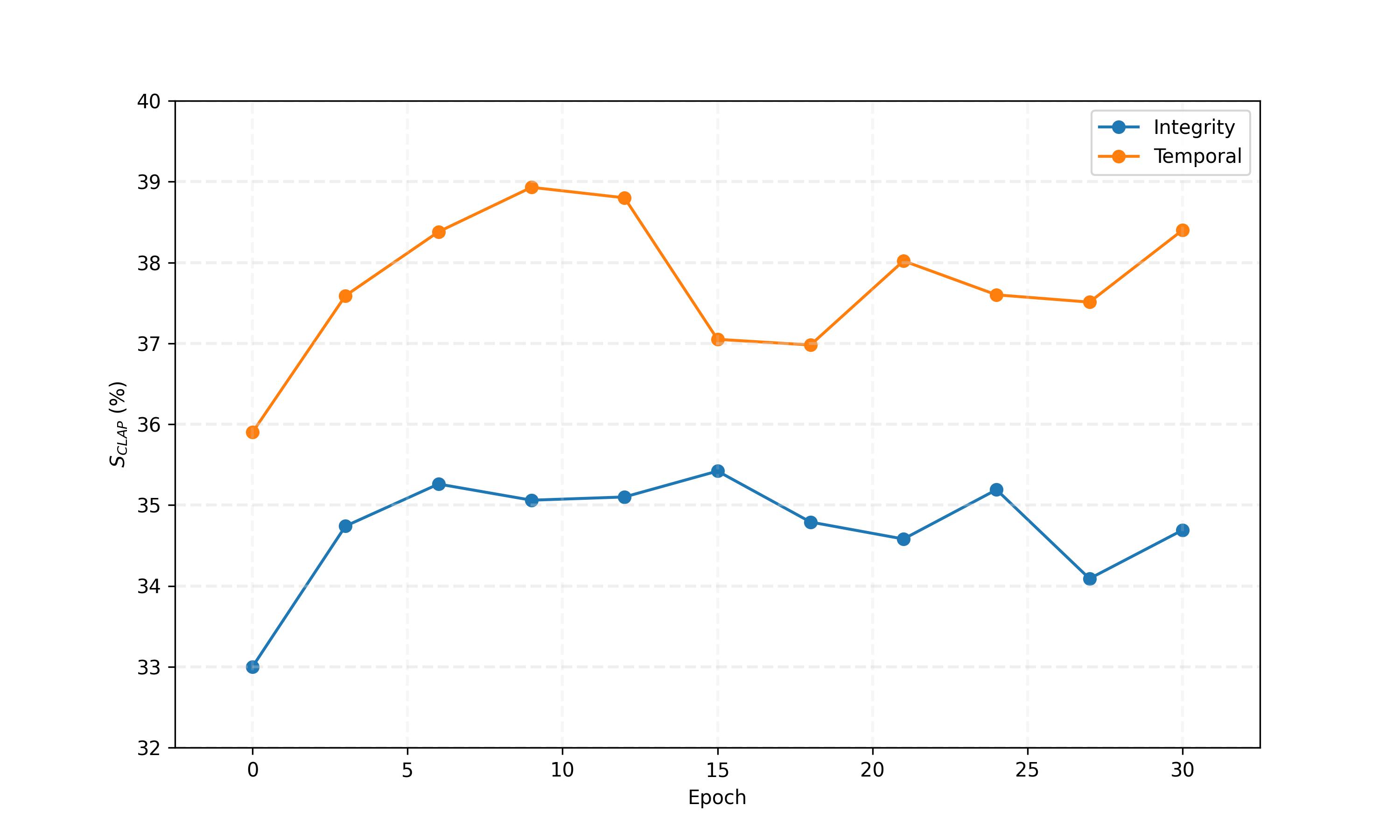}
    \captionsetup{justification=centering}
    \caption{CLAP score variation curve with the fine-tuning iteration.}
    \label{fig:Epoch}
\end{figure}

We increased the number of iterations for fine-tuning, and the results are illustrated in Figure~\ref{fig:Epoch}. In the integrity task, the CLAP score exhibit an initial ascent followed by stabilization with increasing iterations, while for the temporal task, the CLAP score exhibit a rapid ascent followed by a subsequent decline, eventually showing a slight increase again. By the 10th epoch of fine-tuning, the CLAP score for both subtasks reach a higher point and demonstrate greater stability. Hence, the 10th epoch appears to represent a balance point between performance and efficiency, making it a suitable candidate for halting model fine-tuning.

\subsection{Weight parameter $\beta$ for pretrain loss}

Table~\ref{tab:weight} illustrates the impact of pretrain loss weights on both audio quality and alignment. In the integrity task, audio quality shows a gradual improvement with increasing $\beta$, particularly evident in the metric FD. However, there is a concurrent decrease in audio diversity and alignment as $\beta$ increases, which suggests that there is a balance between audio quality and alignment. Regarding the temporal task, comparatively optimal performance is observed when the $\beta$ is set to 0.5, resulting in the highest CLAP score. Additionally, the metric FD, indicative of the quality of generated audio, exhibit favorable performance under this configuration.


\begin{table}[H]
    \centering
    \begin{tabularx}{0.45\textwidth}{l|c|*{3}{X}c|X}
        \toprule
        \small{Task} & \small{$\beta$} & \small{FD$\downarrow$} & \small{FAD$\downarrow$} & \small{IS$\uparrow$} & \small{KL$\downarrow$} & \small{S$_{CLAP}$$\uparrow$} \\
        \midrule
        \multirow{2}{*}{\small{Integrity}} & 0.25 & 40.23 & 2.92 & \textbf{5.20} & 1.15 & \textbf{35.7$\%$} \\
        & 0.5 & 38.54 & 3.02 & 5.07 & 1.11 & 35.3$\%$ \\
        & 0.75 & 38.44 & 3.31 & 4.91 & 1.17 & 33.2$\%$ \\
        & 1 & \textbf{37.78} & \textbf{2.56} & 4.76 & \textbf{1.10} & 34.2$\%$ \\
        \midrule
        \multirow{2}{*}{\small{Temporal}} & 0.25 & 37.23 & 4.68 & 4.92 & 1.12 & 39.4$\%$ \\
        & 0.5 & \textbf{36.19} & 4.31 & 4.88 & 1.13 & \textbf{41.9$\%$} \\
        & 0.75 & 37.37 & \textbf{4.28} & \textbf{5.05} & \textbf{1.09} & 38.7$\%$ \\
        & 1 & 37.76 & 4.29 & 4.72 & 1.14 & 37.7$\%$ \\
        \bottomrule
    \end{tabularx}
    \caption{Impact of weight parameter $\beta$ on audio quality and alignment.}
    \label{tab:weight}
\end{table}

\section{More Samples}

There more samples from original model and our models. The above audio samples can be listened to on our project page. 

\subsection{Integrity Task}

\begin{figure*}[t]
    \centering
    \includegraphics[width=1.06\linewidth]{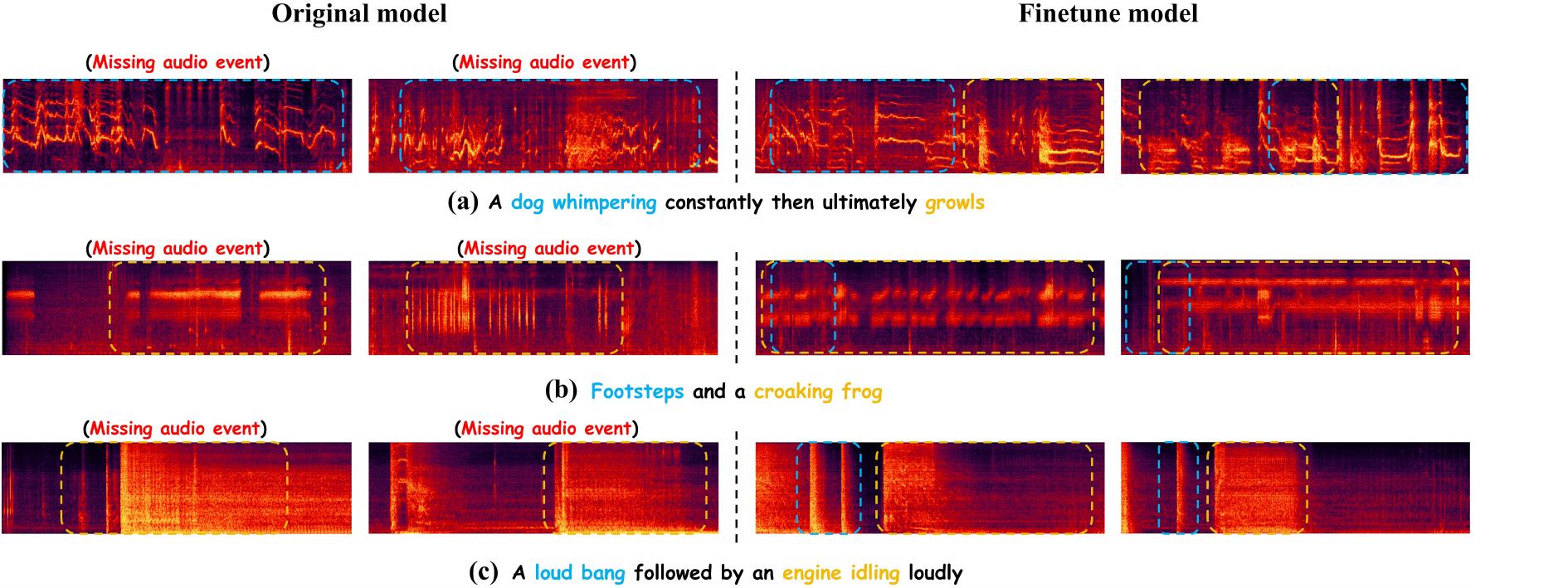}
    \caption{Samples from the integrity task comparing the original model with the fine-tuned model}
    \label{fig:sample1}
\end{figure*}

Figure~\ref{fig:sample1} presents the spectrum of generated audio of both the original and fine-tuned models on the test set for the integrity task. The initial model encountered issues with missing audio event mentioned in the prompt. Following fine-tuning, the occurrence of audio events for the two-label prompt has been enhanced.

\subsection{Temporal Task}

\begin{figure*}[htbp]
    \centering
    \includegraphics[width=1.06\linewidth]{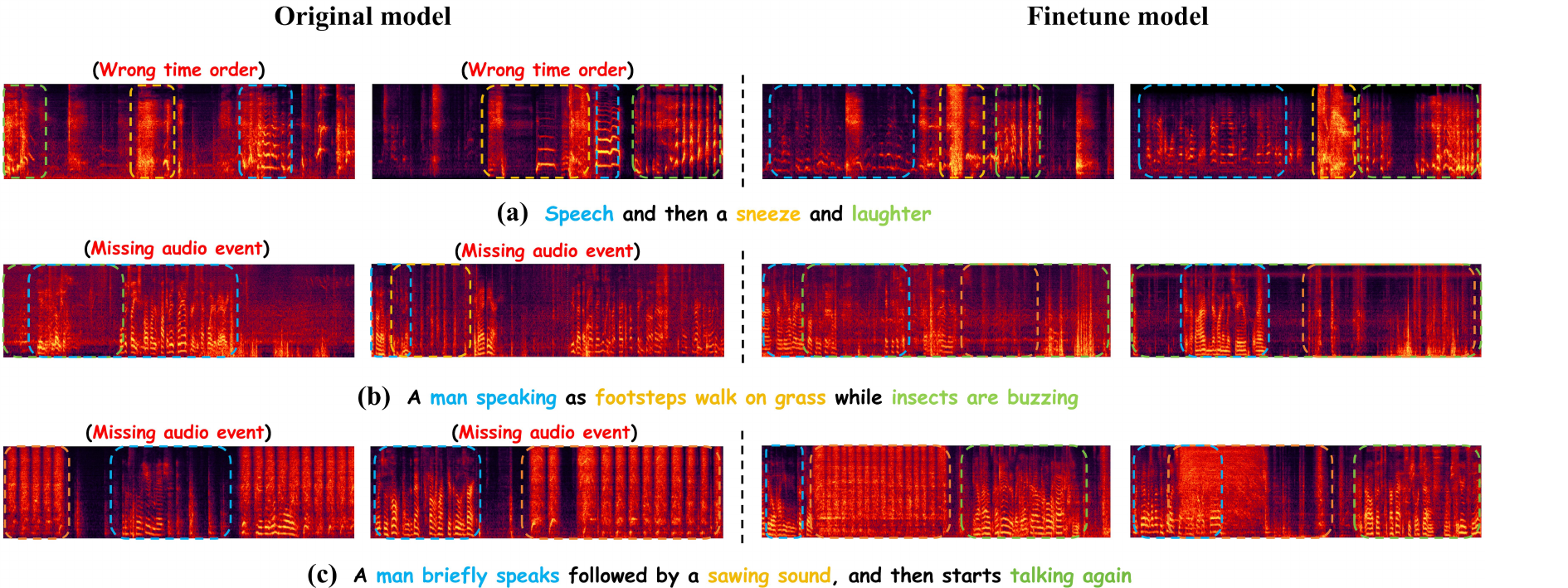}
    \caption{Samples from the temporal task comparing the original model with the fine-tuned model}
    \label{fig:sample2}
\end{figure*}

Figure~\ref{fig:sample2} presents the spectrum of generated audio of both the original and fine-tuned models on the test set for the temporal task. The initial model faced the challenge of generating all the audio without omissions ensuring correct temporal order in accordance with the prompt.

\end{document}